\documentclass[manuscript=article]{achemso}

\usepackage{amsmath}
\usepackage{graphicx}% Include figure files
\usepackage{dcolumn}% Align table columns on decimal point
\usepackage{bm}% bold math
\usepackage{color}
\usepackage{soul}
\usepackage[normalem]{ulem}

\usepackage{pdfpages}

\newcommand{\rmd}{{\rm d}}
\newcommand{\rmi}{{\rm i}}

\author{Jiajun Li}
\affiliation[buffalo]{Department of Physics, State University of New York at Buffalo, Buffalo,
New York 14260, USA}

\author{Camille Aron}
\affiliation[cnrs]{Laboratoire de Physique Th\'eorique, \'Ecole Normale Sup\'erieure,
 CNRS, PSL Research University, Sorbonne Universit\'es, 75005 Paris, France}
\alsoaffiliation[ku]{Instituut voor Theoretische Fysica, KU Leuven, Belgium}

\author{Gabriel Kotliar}
\affiliation[rutgers]{Department of Physics, Rutgers University, New Jersey 08854, USA}

\author{Jong E. Han}
\email{jonghan@buffalo.edu}
\affiliation[buffalo]{Department of Physics, State University of New York at Buffalo, Buffalo,
New York 14260, USA}

\title[]{Microscopic Theory of Resistive Switching in Ordered Insulators:
Electronic vs. Thermal Mechanism}

\begin{document}

%\begin{tocentry}
%\end{tocentry}

\begin{abstract}

We investigate the dramatic switch of resistance in ordered correlated
insulators, when driven out of equilibrium by a strong voltage bias.
Microscopic calculations on a driven-dissipative lattice of interacting
electrons explain the main experimental features of resistive switching
(RS), such as the hysteretic $I$-$V$ curves and the formation of hot
conductive filaments. The energy-resolved electron distribution at the
RS reveals the underlying nonequilibrium electronic mechanism, namely
Landau-Zener tunneling, and also justifies a thermal description where
the hot-electron temperature, estimated from the first moment of the
distribution, matches the equilibrium phase transition temperature. We
discuss the tangled relationship between filament growth and negative
differential resistance, and the influence of crystallographic structure
and disorder in the RS.

\end{abstract}


%\pacs{71.27.+a, 71.30.+h, 72.20.Ht}
%\pacs{}
%\maketitle

\medskip

A variety of correlated oxides~\cite{French, French2, Tokura, Parkin,
inoue, jslee,Doug, noh, shukla} experience a sudden change of
resistivity by several orders of magnitude when subject to strong
external electric fields of $10^{2}\sim 10^{4}$ V/cm.  This
nonequilibrium phase transition, referred as resistive switching (RS), 
shows hysteretic $I$-$V$ characteristics essential for new electronic memory/switching
devices. The switching mechanisms in the RS have been debated for the
last few decades without consensus, and a microscopic understanding of its underlying
mechanism is now essential for continued progress in the field. 
Two underlying mechanisms to RS have been proposed: (1) electronic
mechanism: direct effect of the electric field such as a dielectric breakdown
due to electrons tunneling across the Mott
gap~\cite{oka,tsuji,joura,mazza1,mazza2}, and (2) thermal mechanism:
indirect effect of the electric field with a thermally-induced phase
transition due to Joule heating which locally melts the
insulator~\cite{Chudnovskii, Guenon, Brockman, VO2-Bae,
duchene,VO2-Joule-Basov, CRO-Nakamura,zimmers,singh}.  The latter has
recently received a strong experimental support in Ref.~\cite{zimmers}
where the local temperature could be monitored directly as the sample
was bias-voltage-driven through the RS.  Both these scenarios have been
considered disconnected, and a question central to this intense debate
is whether they are compatible or mutually exclusive.

The RS in vanadium oxides such as vanadium dioxide (VO$_2$) and sesquioxide
(V$_2$O$_3$) have received much theoretical and experimental attention.
 In equilibrium, these prototypical Mott
insulators exhibit a temperature-driven insulator-to-metal phase
transition between an ordered insulator (with dimerized vanadium pairs
in VO$_2$ and antiferromagnetism in V$_2$O$_3$) and a disordered metal
with a resistivity drop of four orders of
magnitude~\cite{Morin59,Takei66}. Out of equilibrium, the RS provoked by
a strong voltage bias is accompanied with the formation of conductive
filaments along the electric field~\cite{berglund, duchene}. These
filaments were interpreted as electrical instabilities related to the
peculiar S-shaped $I$-$V$ characteristics measured in VO$_2$, in
particular to their region of negative differential resistance
(NDR)~\cite{ridley,CopePenn68,hyuntakkim,VO2-JKim}.  

Only a comprehensive microscopic theory of RS in correlated insulators,
compatible with all the experimental evidence at hand, can resolve this
long-standing puzzle.  The recently proposed classical resistor network
models~\cite{French,French2,VO2-Joule-Basov,Frenchreview,driscoll,rozenberg2014,garland}
successfully reproduced part of the phenomenology but these are
heuristic approaches, not firmly grounded from a microscopic
perspective.  In this work, we explain and reproduce the main features
of RS listed above starting from a generic microscopic model which
includes broken symmetry, drive and dissipative mechanisms, and the
spatial inhomogeneity~\cite{potthoff,dobro} for nonequilibrium
phase-segregation.  The quantum calculation further provides important
information on the origin of the nonequilibrium excitations and on how
the electronic and thermal RS scenarios are connected.

Our microscopic description consists of a slab of correlated electrons
of length $L$, a Hubbard model, which is placed between two metallic
leads~\cite{Camille1,Camille2,Amaricci, prl15, graz,satoshi}. A voltage
bias $V_{\rm s}$ across the sample, and the resulting static electric
field $E = V_{\rm s} / L$, is created by connecting the two leads in
series with a resistor $R$ and a dc-voltage generator delivering a total
voltage $V_{\rm t} = V_{\rm s} + R I$, with the current $I$. In addition
to the dissipation by the two leads at the
boundaries~\cite{satoshi,mazza1,rubtsov}, we also introduce energy
relaxation in the bulk~\cite{prl15}.  Both the external resistor and the
energy dissipation are essential modeling ingredients that were
overlooked in previous theoretical approaches of RS. The resistor is
crucial to reveal a non-trivial regime of negative $\rmd I/ \rmd V_{\rm
s}$, and dissipation is crucial to avoid overheating the sample.

We divide the Hamiltonian $\hat{H}$ into (i)  $\hat{H}_{\rm Hub}$, the correlated electronic sample itself, given by a Hubbard model on a finite 2d square lattice, (ii) $\hat{H}_{\rm bath+ leads}$, the two leads and the dissipative environment, given by reservoirs of fermions, and (iii) $\hat{H}_{E}$, the electric-field induced electrostatic potential, originating from our choice to work with the Coulomb gauge~\cite{prl15,han2}.
We have
\begin{equation}
\hat{H}_{\rm Hub} = -t\sum_{\langle
{\boldsymbol r}{\boldsymbol r}'\rangle\sigma}(d^\dag_{\boldsymbol{r}\sigma}d_{\boldsymbol{r'}\sigma}+\mathrm{H.c.})
+ \sum_{{\boldsymbol r}\sigma}\Delta\epsilon_{\boldsymbol r}d^\dagger_{{\boldsymbol r}\sigma}
d_{{\boldsymbol r}\sigma}
 + \, U\sum_{\boldsymbol{r}}\Delta n_{\boldsymbol{r}\uparrow}
\Delta n_{\boldsymbol{r}\downarrow}\,,
\label{hubbard}
\end{equation}
where $d^\dagger_{{\boldsymbol r}\sigma}$  is the fermionic creation operator in the
orbital at site ${\boldsymbol r}$ with spin $\sigma=\uparrow,\downarrow$, and
$\Delta n_{{\boldsymbol r}\sigma}\equiv d^\dagger_{{\boldsymbol r}\sigma}d_{{\boldsymbol
r}\sigma}-1/2$. The hopping integrals given by $t$ are limited to nearest neighbors, while
$U$ controls the on-site Coulombic interaction. To model grain boundaries and defects
in realistic devices, we introduce the possibility of disorder with site-dependent energy levels $\Delta\epsilon_{\boldsymbol r}$.
We set $\hbar=e=a=1$ where $a$ is the lattice spacing and $e$ the electron charge. Below, we work in units of $t$.

The dissipative environment consists of non-interacting fermion
reservoirs coupled to every lattice site~\cite{tsuji,han2,han1,Millis}.
Two non-interacting leads are connected at the sample boundaries.
Reservoirs and leads are in equilibrium at a temperature $T_{\rm
bath}$ that we set to zero unless otherwise stated, we collect them in
\begin{equation}
\hat{H}_{\substack{\text{bath} \\ {}^+\text{leads}}} = \sum_{{\boldsymbol r}\alpha\sigma}\epsilon_\alpha 
c^\dagger_{{\boldsymbol r}\alpha\sigma}c_{{\boldsymbol r}\alpha\sigma}
-\sum_{{\boldsymbol r}\alpha\sigma}{g_{\boldsymbol r}}(d^\dag_{{\boldsymbol
r}\sigma}c_{{\boldsymbol r}\alpha\sigma}+ \mathrm{H.c.})\,.
\end{equation}

The $c^\dagger_{{\boldsymbol r}\sigma\alpha}$'s represent the orbitals
of the reservoirs at sites ${\boldsymbol r}$ in the bulk or at the
boundaries, and $\alpha$ is the continuum index in each reservoir.  The
coupling to the reservoirs is given by $g_{\boldsymbol r}$.  We simply
consider infinite flat bands for the reservoirs dispersions
$\epsilon_\alpha$ and the coupling to the reservoirs yields a
frequency-independent hybridization parameter $\Gamma_{\boldsymbol
r}=\pi |g_{\boldsymbol r}|^2\sum_\alpha \delta(\omega-\epsilon_\alpha)$,
which sets the rate at which particles/energy are exchanged with the
environment.  For the leads, we use $\Gamma_{\boldsymbol r}=\Gamma_{\rm
lead}=1.0$ while the bulk damping rate is set to $\Gamma_{\boldsymbol
r}=\Gamma=0.01$.
The RS is essentially a bulk nonequilibrium phenomenon
and, without the bulk dissipation, the effective temperature in the
steady-state cannot be realistic~\cite{teff} away from the leads.

Finally, with the voltage bias along the $y$-direction, 
the electrostatic potential is given by $\varphi({\boldsymbol r})= - y E$
and
\begin{equation} 
\hat{H}_{E}=\sum_{{\boldsymbol r}\sigma}\varphi({\boldsymbol
r})\left( d^\dagger_{{\boldsymbol r}\sigma}d_{{\boldsymbol r}\sigma} +\sum_\alpha
c^\dagger_{{\boldsymbol r}\alpha\sigma}c_{{\boldsymbol r}\alpha\sigma}\right)\,.
\label{bias}
\end{equation}

To gain insight into the RS, we employ the Hartree-Fock (HF)
approximation to treat the Coulombic interaction in $\hat{H}_{\rm Hub}$.
In equilibrium, this self-consistent mean-field approach produces a
phase transition between  a high-temperature low-$U$ paramagnetic metal
(PM) and a low-temperature large-$U$ antiferromagnetic insulator (AFI).  The corresponding order parameter is the
alternating local
order $\Delta_{\boldsymbol r}$ defined as 
$U\langle n_{{\boldsymbol
r}\sigma}\rangle=(-1)^{n_x+n_y}\Delta_{\boldsymbol r}$
with the lattice coordinates ${\boldsymbol
r}=(n_x,n_y)$.
The retarded and lesser Green's functions for the $d$-orbitals are computed in the steady state via Schwinger-Dyson's equations
$[\boldsymbol{G}_\sigma^{\rm
R}(\omega)]^{-1}_{{\boldsymbol r}{\boldsymbol r}'}=[\omega-\Delta\epsilon_{\boldsymbol
r}-\varphi({\boldsymbol r})-U\langle \Delta n_{{\boldsymbol r},-\sigma}\rangle+\rmi \Gamma_{\boldsymbol r}]\delta_{{\boldsymbol
r}{\boldsymbol r}'}+t\delta_{\langle{\boldsymbol r}{\boldsymbol r}'\rangle}$
and
$
G^{<}_\sigma(\omega)_{{\boldsymbol r}{\boldsymbol r}}=\sum_{\boldsymbol s}|G^{\rm
R}_\sigma(\omega)_{{\boldsymbol r}{\boldsymbol s}}|^2\Sigma^<_{\boldsymbol s}(\omega)
$, respectively. The summation is over all lattice sites ${\boldsymbol s}$ and the lesser
electron self-energy originates from the reservoirs:
$\Sigma^<_{\boldsymbol r}(\omega)=2 \rmi \Gamma_{\boldsymbol
r}f_0(\omega-\mu_{\boldsymbol r})$ with the Fermi-Dirac (FD) distribution
$f_0(\epsilon)=[1+\exp(\epsilon/T_{\rm bath})]^{-1}$ and the local
chemical potential $\mu_{\boldsymbol r}  = \varphi(\boldsymbol r)$.  For
any given voltage $V_{\rm t}$ delivered by the dc-generator, we solve
the problem self-consistently without assuming a specific voltage
profile in the sample~\cite{satoshi,mazza1}.  Starting from an educated
guess, we compute (i) ${G}_\sigma^{\rm R}(\omega)_{{\boldsymbol
r}{\boldsymbol r'}}$, (ii) $G^{<}_\sigma(\omega)_{{\boldsymbol
r}{\boldsymbol r}}$, (iii)  $\langle \Delta n_{{\boldsymbol
r},\sigma}\rangle$ and the total current $I$, (iv) the voltage bias
$V_{\rm s} = V_{\rm t} - R I$ ($R$ is set to 1.2 throughout this work)
and the electric field $E$, and we iterate until convergence is
achieved. We then repeat the procedure by incrementally
changing the total voltage to $V_{\rm t}+\rmd V$ with a small $\rmd V$, and complete the
$I$-$V$ loop. To
prevent the current leak into the fermion baths, we slightly adjust
their chemical potential at each iteration~\cite{suppl}.
While the model is constructed with an AF order, the discussion
can be generalized to other ordered systems.

\begin{figure}
\rotatebox{0}{\resizebox{5.2in}{!}{\includegraphics{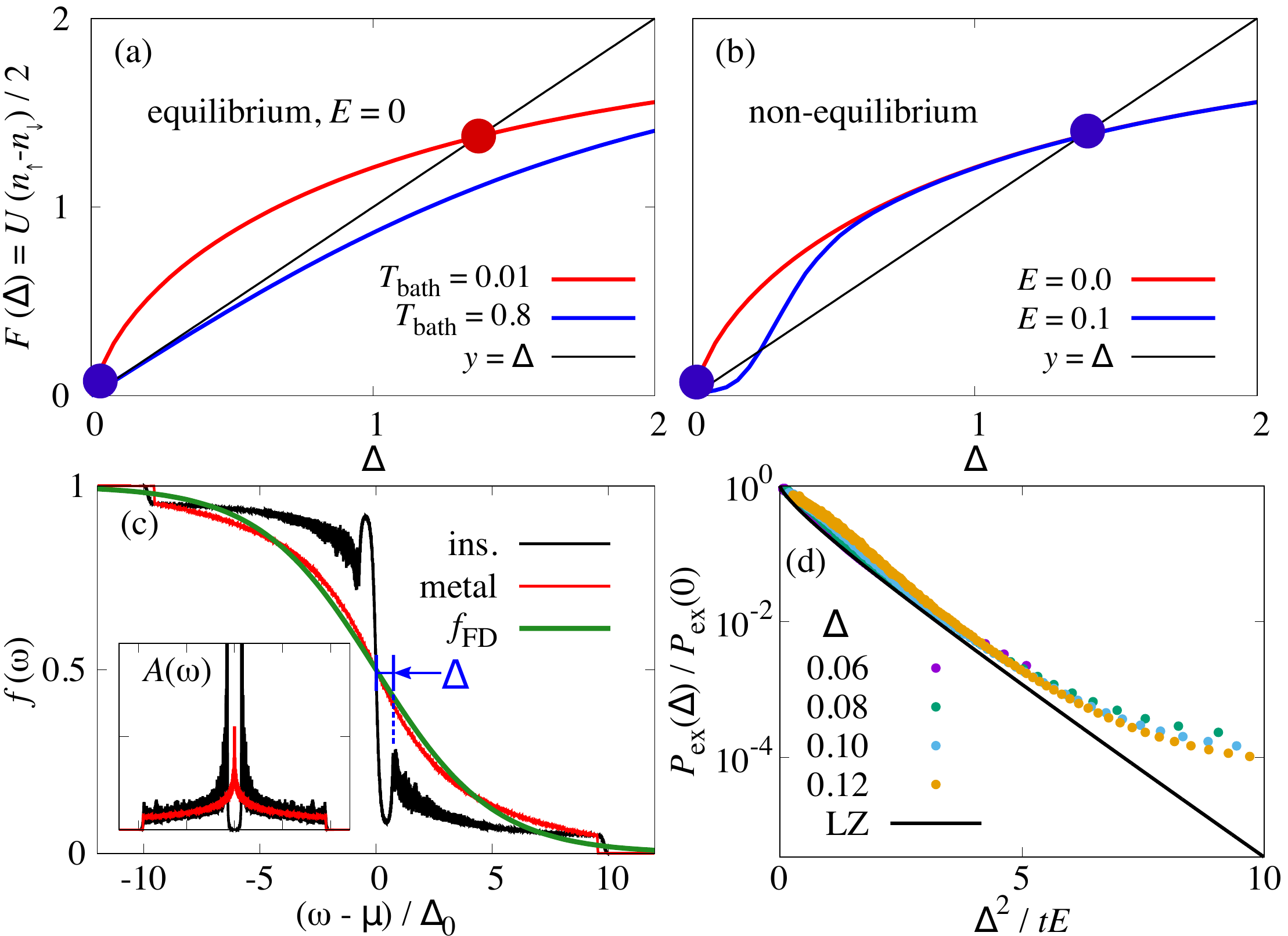}}}
\caption{(a-b) Mean-field condition, $\Delta  = F(\Delta;E,T_{\rm
bath},\Gamma)=\frac12U\langle n_\uparrow-n_\downarrow\rangle$, for a
antiferromagnetic order parameter in an infinite uniform Hubbard model
subject to dissipation.
The stable solutions are marked with dots.
(a) At $E=0$, the antiferromagnetic insulator (AFI) goes continuously to
the paramagnetic metal (PM) $\Delta=0$ upon increasing bath temperature $T_{\rm bath}$.
(b) At finite $E$, Joule heating in the metallic side increases the
effective temperature $T_{\rm eff} > T_{\rm bath}$, stabilizing the PM
solution and resulting in a bi-stable regime. (c) Local energy distribution function
$f(\omega)$ of insulating (black) and metallic (red) phases.
The
Fermi-Dirac function at an effective temperature is shown (green curve,
$T_{\rm eff}=1.05$) for
comparison. Inset: corresponding density of states. (d) Total number of nonequilibrium excitations above the chemical potential, $P_{\rm ex}(\Delta)$. Over
a wide range of gaps,  $P_{\rm ex}(\Delta)$  is well described by the
Fermi-surface averaged Landau-Zener (LZ) tunneling rate~\cite{suppl}
($\Gamma=0.001$).}
\label{resp}
\end{figure}

Let us first discuss the mechanism by which the interplay of drive and
dissipation brings the system to a RS. Within a purely thermal scenario, one
would argue that the nonequilibrium drive effectively enters this
problem only as an effective temperature, $T_{\rm eff}({E})$,
caused by Joule heating.  However, given the strongly discontinuous and
hysteretic RS that are experimentally observed, the underlying mechanism
must go beyond this simple reparametrization of the equilibrium theory.
We address this puzzle by first considering an infinite and
uniform system ($L\to\infty$ keeping $E$ fixed and $\Delta\epsilon_{\boldsymbol r}=0$), where the formalism developed in Ref.~\cite{prl15} can be readily applied to search for uniform steady-state solutions. In equilibrium
($E=0$), the Slater HF theory~\cite{slater} predicts a continuous AFI/PM phase transition at the N\'{e}el temperature $T_{\rm
bath}=T_{\rm N}$. FIG.~\ref{resp}(a) shows the mean-field
conditions on the order parameter $\Delta=F(\Delta;E,T_{\rm
bath},\Gamma)= \frac12 U\langle
n_\uparrow-n_\downarrow\rangle$.  At $T_{\rm bath} <T_{\rm N}$, there is only one
stable AFI solution at a finite $|\Delta| = \Delta_0$ which continuously goes to the PM solution $\Delta=0$
as $T_{\rm bath}>T_{\rm N}$.
The nonequilibrium situation  $(E>0)$ in FIG.~\ref{resp}(b) is dramatically different. There are now two stable solutions~\cite{sugimoto} at low $T_{\rm bath}$: an AFI solution at the equilibrium value $\Delta\approx\Delta_0$ and the PM solution at 
$\Delta = 0$ which was previously unstable.
The latter can be explained within the thermal mechanism with a high effective temperature $T_{\rm eff}\propto E/\Gamma$ caused by Joule heating on the metallic side~\cite{prl15,han2,Millis}, whereas on the insulating side the large gap prevents such an effect.
The intermediate solution is unstable.
The bi-stability of the order parameter results in
heterogeneous phases during the RS, with an
insulator-to-metal transition (IMT) with increasing electric field,
and a metal-to-insulator transition (MIT) with decreasing electric
field.

The mechanism underlying the RS is revealed by the local energy
distribution function $f_{\boldsymbol r}(\omega)=-\frac12{\rm
Im}G^<_\sigma(\omega)_{\boldsymbol{rr}}/{\rm Im}G^{\rm
R}_\sigma(\omega)_{\boldsymbol{rr}}$, see FIG.~\ref{resp}(c). At finite
$E$-fields, the metallic and insulating distributions deviate from the
FD distribution.  Note that, despite the similar overall shape, the
metallic nonequilibrium distribution has a different functional
expression~\cite{Millis} from the FD function (green curve).  In the
insulating phase, there are significant nonequilibrium excitations
beyond the gap $\Delta$.  The total number of nonequilibrium excitations
above the bath chemical potential is plotted in FIG.~\ref{resp}(d) for a
wide range of gaps.  The agreement with the Landau-Zener (LZ) tunneling
rate~\cite{LZ} shows that it is this electronic mechanism which is
responsible for the RS: the electric field accelerates the
quasi-particles in the lower band, which have a finite probability to
tunnel across the gap $\Delta$ and populate the higher band, rendering
the system metallic.  As described above, $\Delta$ is the
self-consistent result of the balance of the electronic interactions,
the non-equilibrium drive and the dissipative mechanisms.  As we shall
see below and in Ref.~\cite{suppl}, this picture is still compatible
with a thermal description where the nonequilibrium excitations are
simply interpreted as thermal excitations. The deviation in the small
$E$ (or large $\Delta^2/tE$) limit is due to the dephasing provided by
the fermion baths.

\begin{figure}
\rotatebox{0}{\resizebox{5.2in}{!}{\includegraphics{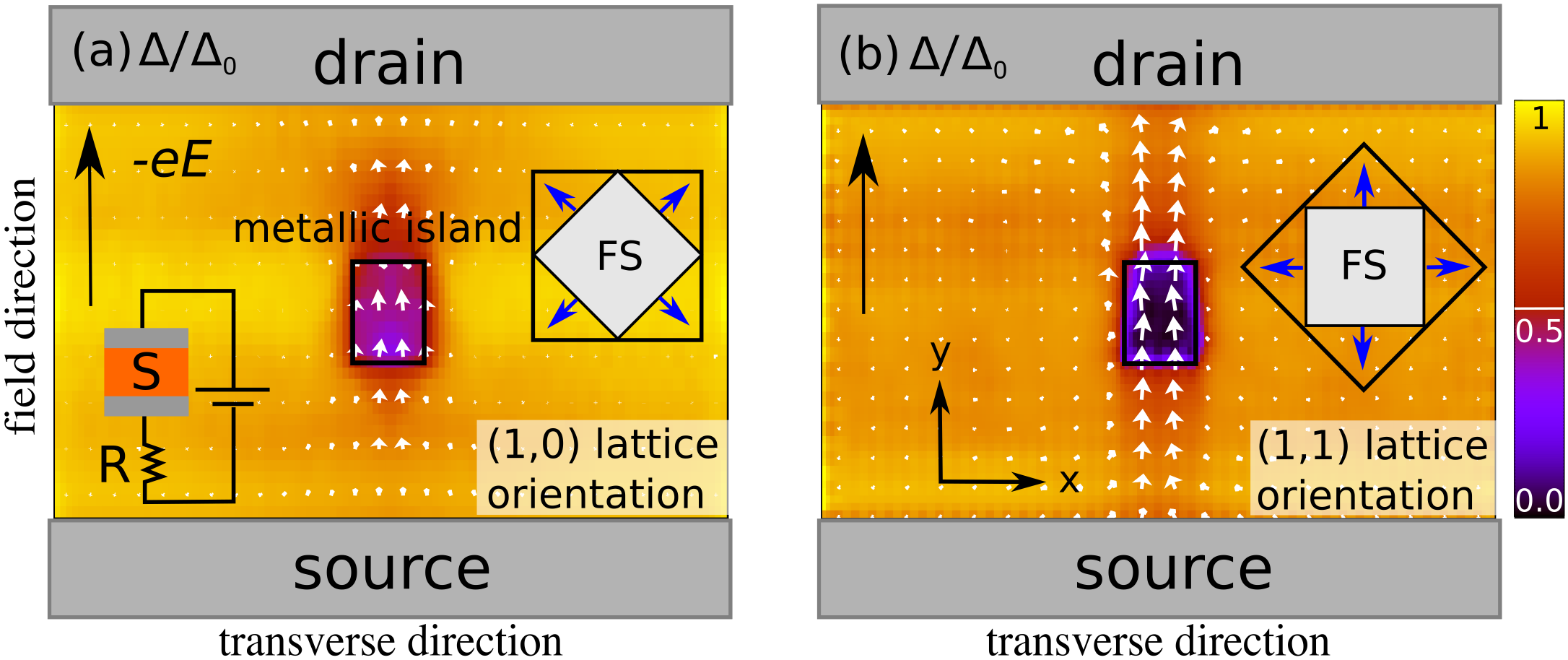}}}
\caption{
Nucleation of a conductive filament in an insulating sample driven by a
voltage bias (see the left inset for a sketch of the setup) and in the
presence of a metallic island.  The color map gives the amplitude of the
antiferromagnetic (AF) order parameter and the white arrows show the
current distribution (direction and amplitude).  (a) Sample cut along
the $(10)$-direction: no filament, the current barely flows outside the
impurity. (b) Sample cut along the $(11)$-diagonal: a robust conductive
filament forms through the impurity, a strong current flows between the
leads.  Right insets: Fermi surfaces and  Fermi-velocity vectors in
blue.  $E/\Delta_0={0.252}$ with the equilibrium gap $\Delta_0={1.35}$,
$U=4.0$ and $T_{\rm bath}=0.3$. 
}
\label{blob}
\end{figure}

Having demonstrated the basic mechanism of the RS, we
now focus on the realistic phenomenology in finite and non-uniform
samples.
As we anticipate that the presence of weak disorder in the form of
impurities or defects may favor the stabilization of mixed phases, we
first investigate the role of spatial inhomogeneities by creating a $5
\times 5$ metallic island (setting $\Delta\epsilon_{\boldsymbol
r}=1.5t$) at the center of an insulating sample with 1200 lattice sites
with size $(80a/\sqrt{2})\times(30a/\sqrt{2})$. In
FIG.~\ref{blob}, we monitor the local order parameter
$\Delta_{\boldsymbol r}$ 
 and the local current
 for two different crystallographic orientations of the square lattice: (a) the sample is cut along the $(10)$-direction, and (b) along the $(11)$-diagonal.
In the former case the RS occurs homogeneously, \textit{i.e.} without noticeable pattern formation, at switching fields close to the values obtained with an infinite and homogeneous lattice.
Yet, in the latter case we found strong and collimated filaments at much weaker fields.

This remarkable anisotropy can be traced to the Fermi surface geometry
of the half-filled square lattice. In a $(11)$-lattice, the
Fermi-velocity vector ${\boldsymbol v}_{\rm F}$ is aligned with the
electric field, see the inset of FIG.~\ref{blob}~(b), making the
diagonal direction an easy-axis for filament formation. This anisotropy
is supported analytically in the non-interacting and weak-field limit,
where the nonequilibrium distribution function can be described by an
anisotropic effective temperature
$T_{\rm eff}(\boldsymbol{E})
\sim \left|{\boldsymbol v}_{\rm F} \cdot
{\boldsymbol E}\right|/\Gamma$.
See Ref.~\cite{suppl} for a detailed discussion. In
polycrystalline samples, the filaments are expected to be globally aligned along $\boldsymbol{E}$, but with domain walls locally aligned along ${\boldsymbol v}_{\rm F}$.

\begin{figure}
\rotatebox{0}{\resizebox{5.2in}{!}{\includegraphics{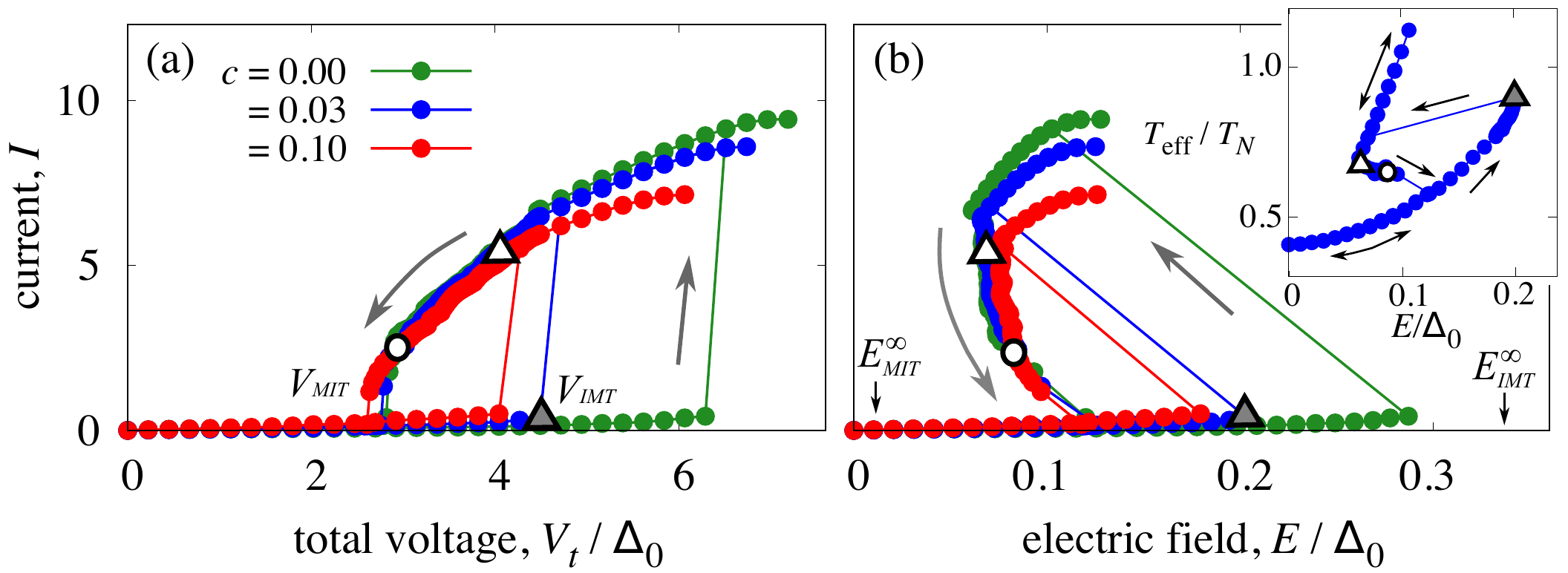}}}
\caption{
$I$-$V$ curves of a rectangular sample cut along the (11)-diagonal for different concentrations $c$ given in the key.
$V_{\rm t}$ is increased then decreased (marked by arrows).
(a) Current as a function of the total voltage $V_{\rm t}$.
Contrary to the MIT, the IMT is strongly affected by the impurity concentration.
(b) Current as a function of the electric field.
When decreasing $V_{\rm t}$, there is a region of negative $\rmd I/\rmd V_{\rm
s}$~\cite{zimmers,hyuntakkim} caused by a narrowing
metallic filament. $E^\infty_{\rm IMT}$ and $E^\infty_{\rm MIT}$
indicate the threshold fields of the infinite uniform lattice. 
In the inset of (b), the averaged effective
temperature at $c=0.03$ displays an hysteretic behavior around the IMT at $T_{\rm eff}\approx T_{\rm N}$.
The upward kink near the white triangle, down the metallic line, is a finite-size effect to disappear in the limit of a large system.
Same parameters as in FIG.~\ref{blob}. 
}
\label{purec}
\end{figure}

\begin{figure}
\rotatebox{0}{\resizebox{5.2in}{!}{\includegraphics{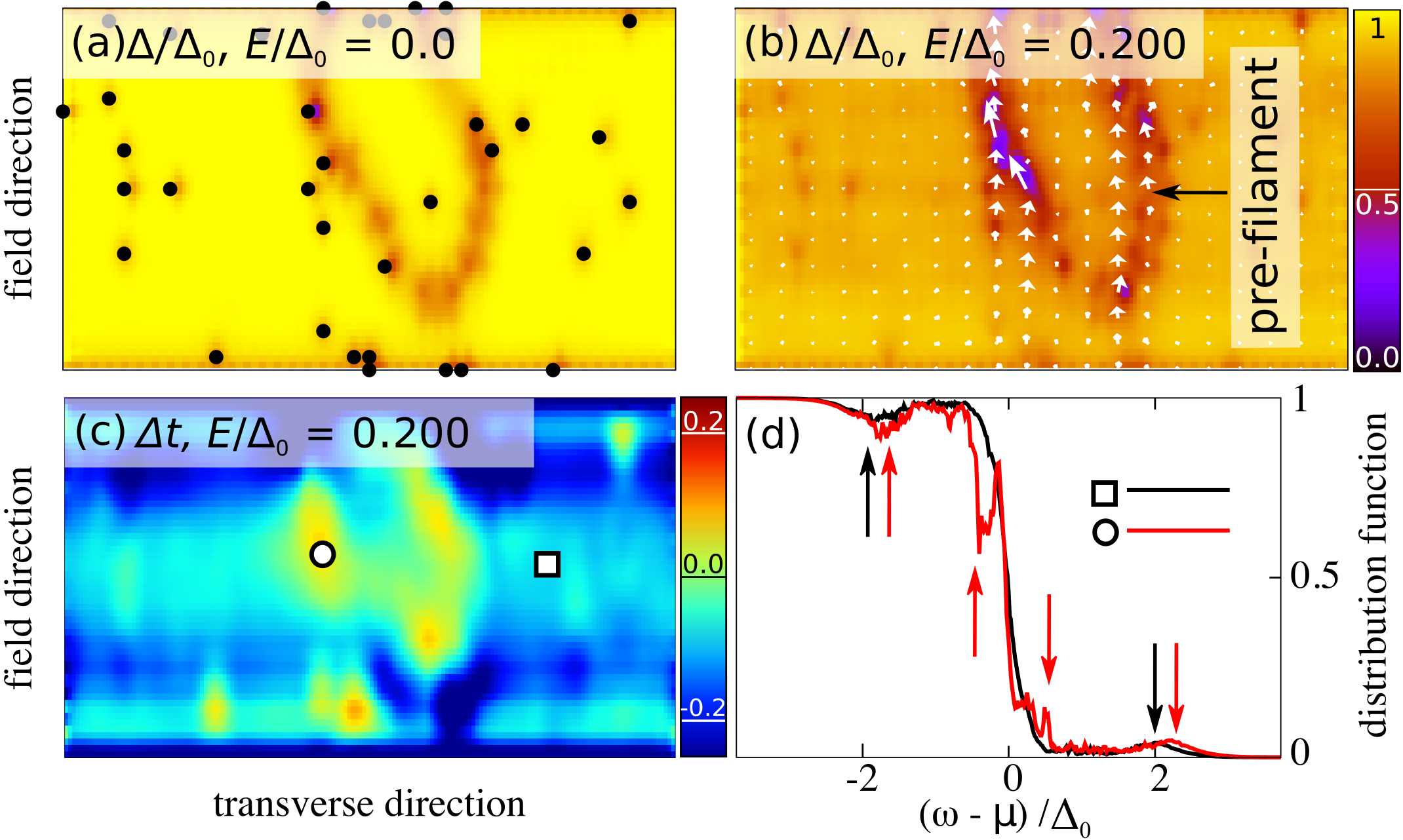}}}
\caption{
Pattern formation ahead of the IMT in a sample cut along the (11)-diagonal.
(a)-(b) AF order parameter (color map) and current (white arrows) for increasing electric field.
(a) At $E=0$, the impurities (black dots) create a unique imprint on the AF order parameter.
(b) At $E/\Delta_0={0.200}$ (grey triangle in Fig.~\ref{purec}), a conductive
path (pre-filament) forms along the low-$\Delta$
region to trigger an IMT. 
(c) Effective temperature $\Delta t \equiv (T_{\rm eff}-T_{\rm N})/T_{\rm N}$
distribution. Note the similarity with the current distribution in (b).
(d) Nonequilibrium distribution function at sites marked in (c). The
arrows point to strong nonequilibrium excitations.
Same parameters as in FIG.~\ref{blob} and $c=0.03$.
}
\label{ipsys}
\end{figure}

We now turn to a model where metallic impurities are
randomly distributed at a fixed concentration $c$. 
FIG.~\ref{purec} shows the hysteretic
behavior of the current for different impurity concentrations as a
function of (a) the total voltage $V_{\rm t}$, and (b) the electric
field $E$. The corresponding RS fields are found to be fractions of the
equilibrium order parameter, $E/\Delta_0 \sim {0.2}$ (see also
FIG.~\ref{ipsys}).  Moreover, the IMT threshold field, $E_{\rm IMT}$,
appears to be strongly reduced in the presence of impurities, while the
MIT at $E_{\rm MIT}$ is barely affected.  This difference is due to the
distinct nature of the two switching mechanisms, as we shall discuss
later.

We now ask whether the nucleation of the filaments at the RS can be interpreted as the result of thermal excitations.
We estimate the local effective temperature $T_{\rm eff}(\boldsymbol{r})$ as a first moment of the distribution function $f_{\boldsymbol r}(\omega) $ via the Sommerfeld-like expansion 
$
\frac{\pi^2}{6}T_{\rm
eff}({\boldsymbol r})^2=\int \! \rmd \omega \, (\omega-\mu_{\boldsymbol
r}) [f_{\boldsymbol r}(\omega)-\theta(\mu_{\boldsymbol r}-\omega)]$.
Averaged over the whole sample, $T_{\rm eff}$ in the inset of
Fig.~\ref{purec}(b) displays an hysteresis in
excellent agreement with the one measured in VO$_2$ in
Ref.~\cite{zimmers}.
In particular, at the IMT we find $T_{\rm eff}  \approx T_{\rm N}
=0.8$, thus
supporting the scenario of a thermally-driven IMT.

The nucleation of a filament at the IMT is a highly nonlinear process
that we discuss in FIG.~\ref{ipsys}. (a) At $E=0$, extended metallic
inhomogeneities (darker zones) lie across the sample, 
and connect the two leads despite the
concentration $c=0.03$ being far below the classical $2d$ percolation
threshold~\cite{percolation}.  The precise pattern of these low-$\Delta$
paths is determined by the impurity distribution but, as a consequence
of a quantum coherence length larger than the impurity spacing, it is
not tightly bound to the impurity locations.  These low-$\Delta$ paths
will act as precursors for the filaments.  At weak fields, they are not
metallic enough to support any linear-response current.  (b)
Only very close
to the IMT at $E/\Delta_0={0.200}$, the filament is
greatly reinforced and now supports a sizeable current.  In (c), we plot
the corresponding effective temperature distribution measured from the
N\'eel temperature, $\Delta t_{\boldsymbol{r}} \equiv [T_{\rm
eff}({\boldsymbol r})-T_{\rm N}]/T_{\rm N}$.  The temperature in the
sample is approximately $T_{\rm N}$, slightly hotter in the pre-filament
region, and slightly cooler close to the leads that are maintained at
$T_{\rm bath}$. As shown in (d), the distribution
function shows hotter region has stronger excitations.

\begin{figure}
\rotatebox{0}{\resizebox{5.2in}{!}{\includegraphics{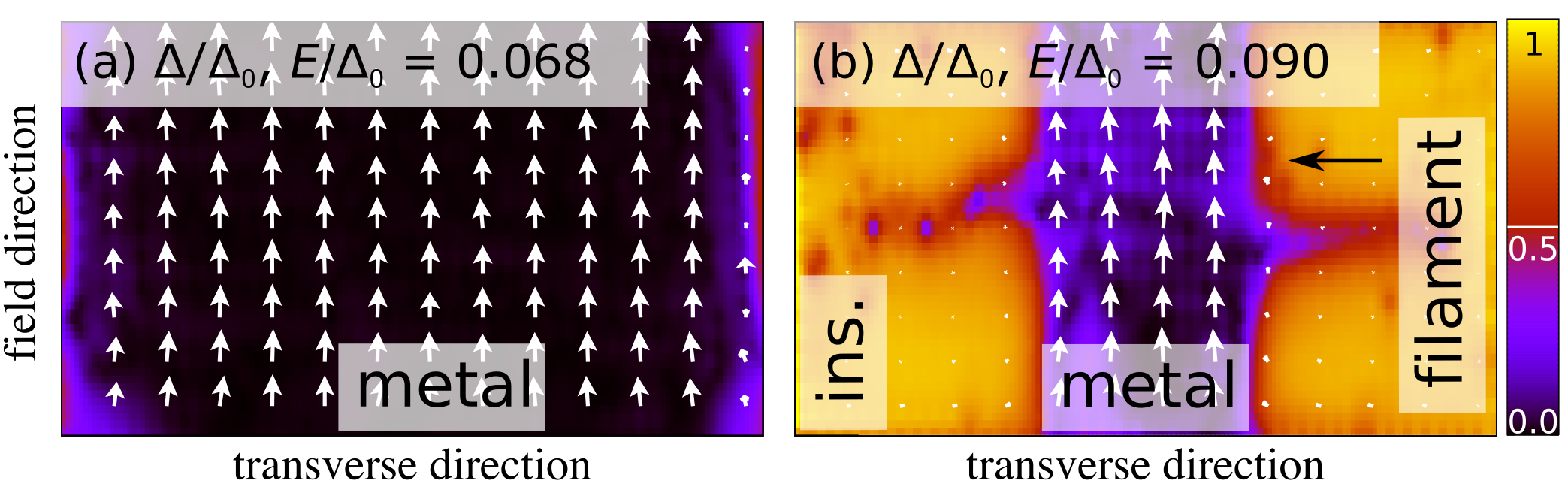}}}
\caption{
Filament formation ahead of the MIT.
(a) At $E/\Delta_0={0.068}$ (white triangle in Fig.~\ref{purec}), the
insulating solution nucleates on the sample edges parallel to the field.
(b) When $\rmd I/ \rmd V_{\rm s} < 0$ [see Fig.~\ref{purec}~(b)], the
insulating domain grows from the edges, leaving a metallic filament
which narrows progressively until the IMT at
$E/\Delta_0={0.090}$  (white circle in Fig.~\ref{purec}). 
}
\label{lform}
\end{figure}

Contrary to the IMT, the MIT is governed by the shrinking of the
filaments upon reducing the bias.  FIG.~\ref{lform}~(a) shows that the
insulating domains start to nucleate from the edges of the sample
parallel to the field at $E/\Delta_0={0.068}$.  This
dependence of the MIT on the sample boundary geometry explains its
rather weak dependence on the bulk impurities. As seen in
FIG.~\ref{lform}~(b), the conducting filament shrinks
as the $E$-field increases, leading to decreasing current.
Remarkably, this filament dynamics results
in the negative differential resistance (NDR) $\rmd I/ \rmd V_{\rm s}<0$
observed in FIG.~\ref{purec}~(b) and also reported in
VO$_2$~\cite{duchene,berglund,zimmers,hyuntakkim}. 
The NDR intrinsically originates from the
nonequilibrium filament dynamics of the ordered solids and this branch of the $I$-$V$
is revealed by adding an external resistor.
(See in Supporting Information~\cite{suppl} for more discussion.)

To summarize, we have identified a minimal quantum driven-dissipative
model that reproduces the main experimental signatures of RS in vanadium
oxides. We showed that the RS is facilitated by a bi-stable
insulator-metal phase which leads to the $I$-$V$ hysteresis, sudden
nucleation of conducting filaments at the IMT, and the progressive
narrowing of the filaments during the NDR regime ahead of the MIT. The
RS  is triggered by a Landau-Zener tunneling process across the
self-consistently determined gap $\Delta$. We furthermore reconciled
this electronic scenario with the Joule heating interpretation by
showing how an effective temperature description could quantitatively
reproduce the amount of excitations in the electronic system.
Investigation by fempto-second STM or photoemission
could resolve the dominant roles in the
electronic and thermal mechanisms proposed in this work.

We are grateful to Petar Bakalov, Keshav Dani, Sambandamurthy Ganapathy,
P\'\i a Homm Jara, Hyun-Tak Kim, Mariela Menghini, Marcelo Rozenberg,
Sujay Singh and  for helpful discussions.  We acknowledge the
computational support at CCR (SUNY at Buffalo).  This work has been
supported by the NSF through the Grant No. DMR-1308141.



\section*{Supplementary material}

Supporting Information:

# Microscopic Theory of Resistive Switching in Ordered Insulators: Electronic vs. Thermal Mechanism

Jiajun Li[1], Camille Aron[2,3], Gabriel Kotliar[4] and Jong E. Han[1,*]

[1] *Department of Physics, State University of New York at Buffalo, Buffalo New York 14260, USA*

[2] *Laboratoire de Physique Théorique, École Normale Supérieure, CNRS PSL Research University, Sorbonne Universités, 75005 Paris, France*

[3] *Instituut voor Theoretische Fysica, KU Leuven, Belgium*

[4] *Department of Physics, Rutgers University, New Jersey 08854, USA*

[*]corresponding author e-mail:`jonghan@buffalo.edu`

## I. CURRENT AND CURRENT LEAK TO THE FERMION RESERVOIRS

The steady-state electric current between nearest neighbor sites $\boldsymbol{r}$ and $\boldsymbol{r}+\hat{\boldsymbol{e}}$ is evaluated as

$$I_{\boldsymbol{r},\hat{\boldsymbol{e}}} = -\mathrm{i}t\sum_\sigma(\langle d^\dagger_{\boldsymbol{r}+\hat{\boldsymbol{e}}\sigma}d_{\boldsymbol{r}\sigma}\rangle - \mathrm{H.c.}) = -t\sum_\sigma\int\left[G^<_{\boldsymbol{r}+\hat{\boldsymbol{e}},\boldsymbol{r}}(\omega) - G^<_{\boldsymbol{r},\boldsymbol{r}+\hat{\boldsymbol{e}}}(\omega)\right]\frac{\mathrm{d}\omega}{2\pi}. \tag{1}$$

where the lesser Green's functions $G^<_{\boldsymbol{rr'}}(\omega)$ are expressed in Eq. (7).

While the role the two leads is to supply both particle and heat transfer to/from the sample, the purpose of the local fermion reservoirs is to provide energy relaxation only. A uniform steady-state current in an infinite homogeneous lattice does not yield any particle flux to the fermion reservoirs that constitute the thermostats when their chemical potential is set to $\mu_{\boldsymbol{r}} = \varphi(\boldsymbol{r})$ [3]. However, in a finite or disordered lattice with a non-uniform current there can be a small volumic leakage of particles to the reservoirs, which can be expressed as

$$I_{\mathrm{leak},\sigma\boldsymbol{r}} = -\mathrm{i}\sum_{\alpha\sigma}g_{\boldsymbol{r}}(\langle d^\dagger_{\boldsymbol{r}\sigma}c_{\boldsymbol{r}\alpha\sigma}\rangle - \mathrm{H.c.}) = 2\Gamma_{\boldsymbol{r}}\int\mathrm{d}\omega\,A_{\sigma\boldsymbol{r}}(\omega)[f_{\sigma\boldsymbol{r}}(\omega) - f_0(\omega - \mu_{\boldsymbol{r}})], \tag{2}$$

with the local density of states $A_{\sigma\boldsymbol{r}}(\omega) = -\pi^{-1}\mathrm{Im}\,G^{\mathrm{R}}_\sigma(\omega)_{\boldsymbol{rr}}$ and the local nonequilibrium distribution function $f_{\sigma\boldsymbol{r}}(\omega) = G^<_\sigma(\omega)_{\boldsymbol{rr}}/(2\pi\mathrm{i}A_{\sigma\boldsymbol{r}}(\omega))$. Therefore, we slightly adjust the local chemical potential $\mu_{\boldsymbol{r}}$ at each HF iteration to ensure the vanishing of $I_{\mathrm{leak},\sigma\boldsymbol{r}}$ and prevent any particle leak.

## II. NEGATIVE DIFFERENTIAL RESISTANCE SOLUTION VIA EXTERNAL RESISTOR

In this Section, we explain the relations between $I(V_\mathrm{t})$ in FIG. 3(a) and $I(V_\mathrm{s})$ in FIG. 3(b) of the Letter, and how to access the negative differential resistance (NDR) regime of the latter.

In FIG. 3(a), the current is plotted as a function of the total voltage imposed by the dc generator on the circuit, $V_\mathrm{t}$. $I(V_\mathrm{t})$ corresponds to what is directly measured in experiments. However, in order to characterize a sample independently of the details of the external electrical circuit, one is interested in giving the current versus the voltage drop across the sample, $V_\mathrm{s}$. $I(V_\mathrm{s})$ is the *intrinsic* characteristic of the sample, plotted in FIG. 3(b).

While both $I(V_\mathrm{s})$ and $I(V_\mathrm{t})$ are of course trivially related to one-another by Kirchhoff's laws, which in our case simply yield

$$V_\mathrm{t} = V_\mathrm{s} + RI, \tag{3}$$

the measurement of $I(V_\mathrm{t})$ for all $V_\mathrm{t}$ does not guaranty the full knowledge of $I(V_\mathrm{s})$. Indeed, if the latter is non-monotonous or multivalued, parts of the $I(V_\mathrm{s})$ characteristic will be inaccessible, *i.e.* hidden, unless the external resistance is selected with care [5].

To illustrate this point, let us re-write Eq. (3) as

$$\frac{V_\mathrm{t} - V_\mathrm{s}}{R} = I(V_\mathrm{s}). \tag{4}$$

This expresses that fact that for a given $V_\mathrm{t}$ delivered by the dc generator, both $I$ and $V_\mathrm{s}$ are self-consistent solutions that live at the intersection(s) of the line $(V_\mathrm{t} - V_\mathrm{s})/R$ (LHS) with the characteristic $I(V_\mathrm{s})$ (RHS).

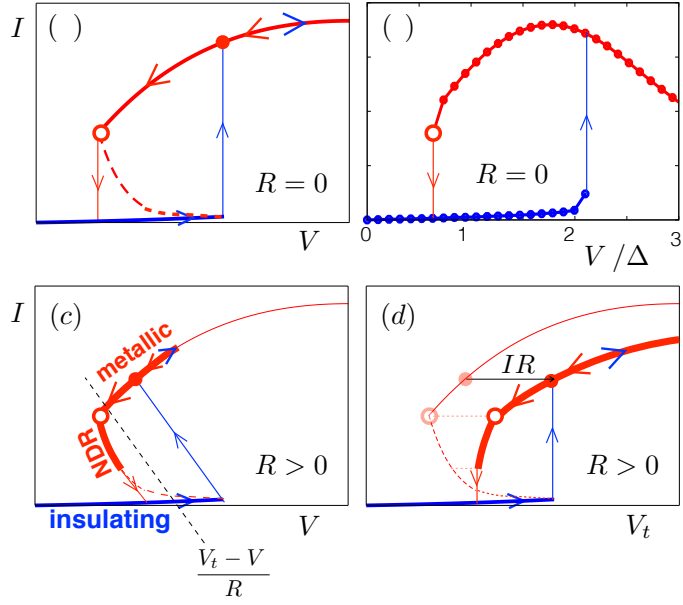

Figure S1: (a) Schematic $I$-$V$ plot with zero external resistance $R = 0$ and (b) the actual numerical results ($c = 0.03$). Schematic graphs (c) plotted against the sample voltage $V_s$ with $R > 0$ and (d) plotted against the total voltage $V_t$ with $R > 0$. The blue (resp. red) arrows denote the solutions obtained when increasing (resp. decreasing) the total voltage $V_t$. White circles mark the first appearance of insulating regions on the edges of the metallic sample when decreasing $V_t$ on the metallic branch.

In our case, we are dealing with S-shaped $I(V_s)$ characteristics, for which the current happens to be multivalued for given values of the voltage, see the sketch in FIG. S1(a). Another case, which we shall not examine here, are N-shaped $I(V_s)$ characteristics for which it is the voltage that is multivalued for a given current. In our case, the physical reason behind that peculiar S shape is the appearance (see the white circles in FIG. S1) and the rapid growth of insulating domains when reducing the voltage bias ahead of the MIT, as shown in FIG. 5(a-b) of manuscript. This causes an NDR regime, $dI/dV_s < 0$, which extends until the sample is fully metallic. For the discussion, we decompose the $I(V_s)$ characteristic in three branches: the insulating branch, the metallic branch, and the NDR branch in the middle, see FIG. S1(a).

To understand why the NDR branch of the $I(V_s)$ characteristic needs extra care to be revealed, let us first consider the case $R = 0$ illustrated schematically in FIG. S1(a) and computed numerically in FIG. S1(b). At $V_t = 0$, there is a single trivial solution: $I = 0$ and $V_s = 0$. As $V_t$ is slowly increased, the solution remains by continuity on the insulating branch of the $I(V_s)$ characteristic, even when two extra solutions have appeared on the other branches of the S. For larger voltages, when $V_t = V_{\text{IMT}}$, the insulating branch stops, and the current makes a brutal and discontinuous transition towards the solution on the metallic branch. As $V_t$ is now slowly decreased, the systems remains on the metallic branch, until it stops at $V_t = V_{\text{MIT}}$, where the current jumps back on the insulating branch. We see that without external resistance, $R = 0$, the system experiences a strong hysteresis making it impossible to probe the NDR branch of the $I(V_s)$ characteristic, no matter the value of $V_t$

It is not hard to show that the condition on the resistance to avoid any hysteresis loop, and

3

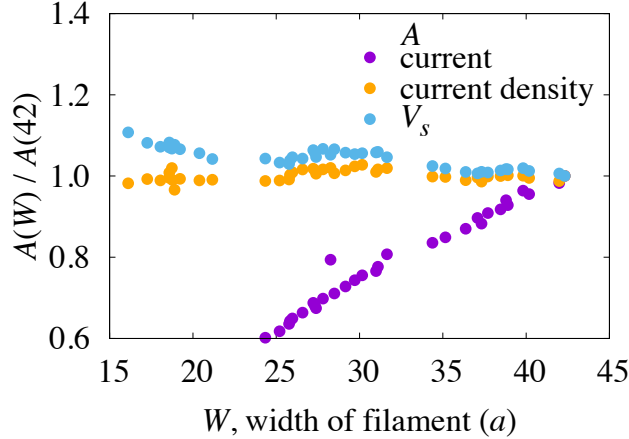

Figure S2: Current and sample voltage $V_s$ as a function of the filament width $W$ on the negative-differential-conductance branch. Inside the filament, the electric current density (measured and averaged alng the central line of the filament) is nearly independent of $W$ while the voltage slightly increases as the width decreases.

therefore probe the full NDR branch, is

$$R > R_{\min} \text{ with } \frac{1}{R_{\min}} \equiv \min\left\{\left|\frac{dI(V_s)}{dV_s}\right|; V_s \in [V_{\text{MIT}}, V_{\text{IMT}}]\right\}, \quad (5)$$

which guaranties a unique solution to Eq. (4) for all values of $V_t$. This is particularly clear in the limit $R \to \infty$, where the LHS of Eq. (4) is an horizontal line, always intersecting $I(V_s)$ at a single point. Such a current-controlled measurement has been performed experimentally in Ref. [6], where the NDR branch was thereby fully uncovered.

In the intermediate cases, $0 < R < R_{\min}$, the NDR branch can only be partially revealed, as shown in FIG. S1(c). This is the case for the numerical data presented in the manuscript (the external resistance was set to $R = 0.634$).

## III. NEGATIVE DIFFERENTIAL RESISTANCE MECHANISM

The NDR in the resistive switching is characterized by the metal-insulator coexistent phase. The main experimental observation is that the $I$-$V$ curve is very steep [6, 7] with the slope $|dI/dV_s|$ decreasing with increasing smaple bias $V_s$, which was reproduced by the calculation in the main text. The mechanism for the NDR can be understood as follows. As shown in Figure S2, the current density inside the filament is the property of the bulk metallic state and remains nearly constant as a function of the filament width $W$. Therefore, the total current is linearly proportional to $W$, as clearly shown in the figure. On the other hand, the voltage drop in the sample is given by the *resistivity* of the metallic state and current density, with the *resistance* being nearly independent of the current, leading to a vertical $I$-$V$ relation. However, as indicated by the data in the figure, the voltage drop slightly increases at small $W$, due to the enhanced scattering from domain boundaries of the filament. This opposite behavior of $I$ and $V$ as a function of the filament width is the origin of the NDR. For instance, an increase of current is mainly due to the growing filament width, which results in less electron scattering and reduced resistance and voltage drop across the sample.

4

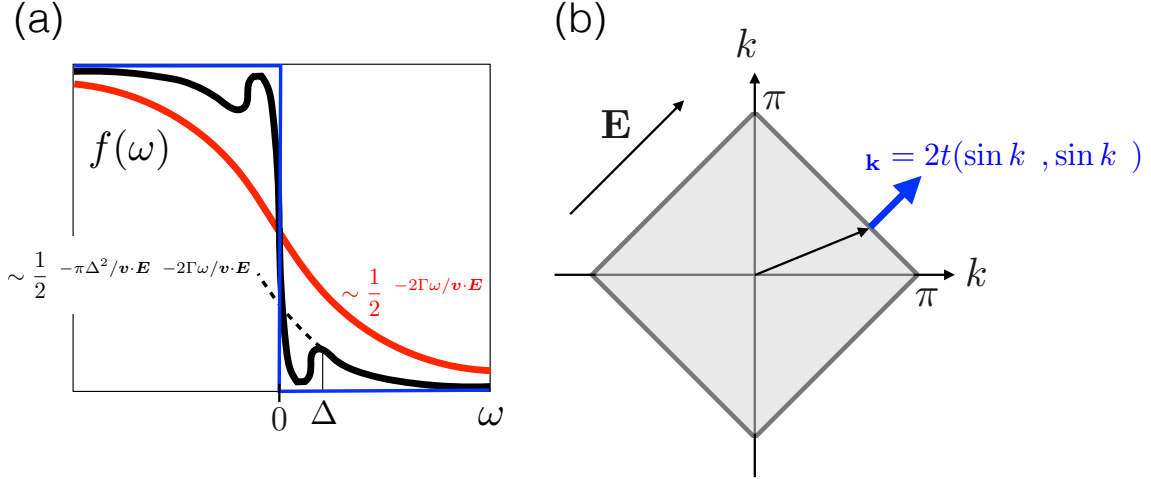

Figure S3: (a) Schematic plot of the local energy distribution function given in FIG. 1 of the Letter. The metallic distribution (red), with a tail $\sim \frac{1}{2}e^{-\Gamma\omega/vE}$, is reduced in the gapped phase (black) by the Landau-Zener factor $e^{-\pi\Delta^2/\boldsymbol{v}\cdot\boldsymbol{E}}$ as outlined by the dashed line that fits most of the nonequilibrium energy distribution. The equilibrium distribution (blue) at $E = 0$ is a step function because the environment is kept at zero temperature. (b) Fermi sea of a $2d$ square lattice at half-filling. The electric field $\mathbf{E}$ is oriented along the $(1, 1)$ direction, as in FIG. 1 of the Letter.

This demonstrates that the NDR in the RS is an *intrinsic* property of the sample, independent of the external resistor. The external resistor only helps the algorithm to find the NDR branch out of the three possible stable solutions for $E$-field with $E_{\text{MIT}} < E < E_{\text{IMT}}$. We can numerically prove that the NDR is an intrinsic property of the sample, as follows. We proceed as in Figure S1(c) from the metallic branch by decreasing the total voltage $V_t$, and stop the simulation as soon as the NDR solution is found, right past the white circle. We then remove the external resistor from the circuit and increase the voltage which is now the sample voltage $V_s$. This procedure reproduces the same NDR $I$-$V$ relation as shown in Figure S1(c). This behavior was confirmed on the curves reported on the main text.

## IV. ENERGY DISTRIBUTION FUNCTION

In this Section, we discuss the local energy distribution function which provides valuable quantum mechanical information about the nonequilibrium dynamics. It is defined as

$$f_{\boldsymbol{r}}(\omega) = -\frac{1}{2}\frac{\text{Im}G^<_\sigma(\omega)_{\boldsymbol{rr}}}{\text{Im}G^R_\sigma(\omega)_{\boldsymbol{rr}}}. \quad (6)$$

and reduces to the Fermi-Dirac function in equilibrium ($E = 0$). Within the HF approximation, the lesser Green's function can be expressed as [3]

$$G^<_{\boldsymbol{rr}'}(\omega) = 2i\Gamma \sum_{\boldsymbol{r}''} G^R_{\boldsymbol{rr}''}(\omega)[G^R_{\boldsymbol{r}''\boldsymbol{r}'}(\omega)]^* f_0(\omega + \boldsymbol{r}'' \cdot \boldsymbol{E}), \quad (7)$$

and locally, we have

$$G^<_{\boldsymbol{00}}(\omega) = 2i\Gamma \sum_{\boldsymbol{r}} |G^R_{\boldsymbol{0r}}(\omega)|^2 f_0(\omega + \boldsymbol{r} \cdot \boldsymbol{E}), \quad (8)$$

5

where $f_0(\omega)$ is the Fermi-Dirac distribution of the fermion reservoirs maintained in equilibrium at temperature $T_{\text{bath}}$. In the case of a pure metal (non-interacting limit), it was already shown in Ref. [3] that the precise shape of the nonequilibrium distribution function is the result of superimposing the Fermi-Dirac distributions originating from the fermion reservoirs at distant sites with chemical potential governed by the bias potential $\varphi(\boldsymbol{r})$. It is the overall profile of the resulting nonequilibrium distribution that can be used to define an effective temperature $T_{\text{eff}}$, *a priori* much different from $T_{\text{bath}}$. See, for example, FIG. S2(a) where the driven metallic phase (in red) displays a hot $T_{\text{eff}}$. We shall discuss this effective temperature in more details in Sect. V. Noteworthy, one can interpret the above factor $|G^R_{\boldsymbol{0r}}(\omega)|^2$, which is the wavefunction overlap between sites $\boldsymbol{r}$ and $\boldsymbol{0}$ due to quantum tunneling, as the coupling strength of the statistical information across the lattice. We emphasize that this nonequilibrium distribution is the result of electronic processes only. Therefore, it should establish itself much faster than a simple thermal diffusion process.

The above formal discussion on the electronic origin of the nonequilibrium distribution function can be corroborated by showing explicitly that the nonequilibrium excitations correspond to Landau-Zener tunneling processes (LZT) [9] over the gap $\Delta$. If this claim is correct, it implies that the nonequilibrium distribution functions of the metallic phase (where $\Delta = 0$) and the insulating phase (where $\Delta > 0$) mostly differ by a LZT factor $e^{-\pi\Delta^2/|\boldsymbol{v}\cdot\boldsymbol{E}|}$, as depicted schematically in FIG. S2(a). To test this claim, we compare the numerical results with the LZT predictions on the total number of nonequilibrium excitations above the chemical potential,

$$P_{\text{ex}}(\Delta) = \int_0^\infty A(\omega;\Delta) f(\omega;\Delta) d\omega, \tag{9}$$

where $A(\omega) = -\pi^{-1} \text{Im} G^R_{\text{loc}}(\omega)$ is the local density of states. On the one hand, we compute Eq. (9) numerically. On the other hand, we estimate $P_{\text{ex}}(\Delta)$ as the steady-state number of excitations in the presence of two competing processes: LZT events occurring at a rate $\gamma_{\text{LZ}}(\Delta) \sim E \exp\left(-\pi\Delta^2/|\boldsymbol{v_k}\cdot\boldsymbol{E}|\right)$ (the precise numerical prefactor arises from the Bloch oscillations frequency), and relaxation events occurring at a rate $\sim \Gamma$. We therefore obtain, summed over the Fermi surface (FS),

$$P_{\text{ex}}(\Delta) = \frac{1}{\mathcal{S}_{\text{FS}}} \int_{\boldsymbol{k}\in \text{FS}} d\boldsymbol{k}\, \frac{\gamma_{LZ}(\Delta)}{\gamma_{LZ}(\Delta)+\Gamma}, \tag{10}$$

where $\mathcal{S}_{\text{FS}}$ is the Fermi surface area. In our case of a $2d$ square lattice with the field $\boldsymbol{E}$ along the $(1,1)$ diagonal, the group velocity at the half-filled Fermi surface is either parallel or perpendicular to $\boldsymbol{E}$. We can therefore reduce the Fermi surface integral to its first quadrant defined as $k_x + k_y = \pi$ [see Fig. S2(b)], yielding the prediction

$$P_{\text{ex}}(\Delta) = \frac{1}{\pi}\int_0^\pi dk_x \frac{\frac{E}{\sqrt{2}}\exp\left(-\frac{\pi\Delta^2}{2\sqrt{2}tE\sin k_x}\right)}{\frac{E}{\sqrt{2}}\exp\left(-\frac{\pi\Delta^2}{2\sqrt{2}tE\sin k_x}\right)+\Gamma}. \tag{11}$$

In FIG. 1(d) of the Letter, we compare this LZT prediction with the numerics. They display an excellent agreement over several orders of magnitude of the gap $\Delta$. This demonstrates that the Landau-Zener tunneling is responsible for the nonequilibrium excitations.

6

## V. EFFECTIVE TEMPERATURE

In this Section, we discuss and justify the expression for the effective temperature

$$T_{\text{eff}} \sim \frac{|\boldsymbol{E} \cdot \boldsymbol{v}_{\text{F}}|}{\Gamma}, \quad (12)$$

which we use inside metallic domains to argue for the strong dependence of Joule heating on the crystallographic direction with respect to the field direction. Aside from this geometric consideration, note that this expression also transparently elucidated how Joule heating sets the temperature as the result of a balance between the drive $\boldsymbol{E}$ and the energy relaxation rate $\Gamma$.

In the metallic regime, we can start by neglecting the on-site Coulomb interaction. Within a first-order gradient approximation, *i.e.* when the spatial and temporal inhomogeneities of the system occur on mesoscopic scales, the Quantum Boltzmann Equation governing the distribution function $f_\sigma$ reads, dropping the spin index,

$$[\partial_T + \boldsymbol{v}(\boldsymbol{k}) \cdot \boldsymbol{\nabla}_{\boldsymbol{X}} + \boldsymbol{E} \cdot \boldsymbol{\nabla}_{\boldsymbol{k}}] f(X; k) = 2\Gamma \left[ f_0(\omega - \mu_{\boldsymbol{X}}) - f(X; k) \right]. \quad (13)$$

Here, $X = (T, \boldsymbol{X})$ are space-time coordinates and $k = (\omega, \boldsymbol{k})$ are Fourier components. The velocity is given by $\boldsymbol{v}(\boldsymbol{k}) = \boldsymbol{\nabla}_{\boldsymbol{k}} \epsilon(\boldsymbol{k})$ where $\epsilon(\boldsymbol{k})$ is the dispersion relation (square lattice in our case). Having neglected the Coulomb interactions, the collision integral on the RHS is only due to the scattering with the degrees of freedom of the local thermostats in equilibrium at temperature $T_{\text{bath}}$ and chemical potential $\mu_{\boldsymbol{X}} = -\boldsymbol{E} \cdot \boldsymbol{X}$. $f_0(\epsilon) \equiv [1 + \exp(\epsilon/T_{\text{bath}})]^{-1}$ is the Fermi-Dirac distribution. Once a steady state is reached, it simplifies to

$$[\boldsymbol{v}(\boldsymbol{k}) \cdot \boldsymbol{\nabla}_{\boldsymbol{X}} + \boldsymbol{E} \cdot \boldsymbol{\nabla}_{\boldsymbol{k}}] f(X; k) = 2\Gamma \left[ f_0(\omega + \boldsymbol{E} \cdot \boldsymbol{X}) - f(X; k) \right]. \quad (14)$$

If analyzing this equation order by order in $\boldsymbol{E}$, *i.e.* decomposing $f(\boldsymbol{X}; k) = f_0(\omega) + \boldsymbol{E} \cdot \boldsymbol{f}^{(1)} + E_i E_j f_{ij}^{(2)} + \ldots$, one can check that the contribution from the term in $\boldsymbol{E} \cdot \boldsymbol{\nabla}_{\boldsymbol{k}}$ in Eq. (14) is of higher order than those of the other terms and can therefore be neglected in a first approximation. Therefore we simplify further the equation to work with

$$\boldsymbol{v}(\boldsymbol{k}) \cdot \boldsymbol{\nabla}_{\boldsymbol{X}} f(X; k) = 2\Gamma \left[ f_0(\omega + \boldsymbol{E} \cdot \boldsymbol{X}) - f(X; k) \right]. \quad (15)$$

In the limit of zero-temperature thermostats, $T_{\text{bath}} \to 0$, we obtain the following solution

$$f(\boldsymbol{X}; \omega, \boldsymbol{k}) = \Theta(-\omega - \boldsymbol{E} \cdot \boldsymbol{X})$$
$$+ \frac{1}{2} \left[ \text{sign}(\omega + \boldsymbol{E} \cdot \boldsymbol{X}) + \text{sign}(\boldsymbol{E} \cdot \boldsymbol{v}(\boldsymbol{k})) \right] \exp \left( -2\Gamma \left| \frac{\omega + \boldsymbol{E} \cdot \boldsymbol{X}}{\boldsymbol{E} \cdot \boldsymbol{v}(\boldsymbol{k})} \right| \right). \quad (16)$$

where $\Theta(\epsilon)$ is the Heaviside step function. For wave-vectors $\boldsymbol{k}^*$ such that $\boldsymbol{E} \perp \boldsymbol{v}(\boldsymbol{k}^*)$, this simplifies to

$$f(\boldsymbol{X}; \omega, \boldsymbol{k}^*) = \Theta(-\omega - \boldsymbol{E} \cdot \boldsymbol{X}), \quad (17)$$

which is the zero-temperature Fermi-Dirac distribution, $\lim_{T_{\text{bath}} \to 0} f_0(\omega - \mu_{\boldsymbol{X}})$. On the contrary, for wave vectors such that $\boldsymbol{v}(\boldsymbol{k})$ is parallel to the local $\boldsymbol{E}$ field, the distribution is far from the zero-temperature equilibrium of the thermostats. One can extract an effective temperature from the solution expressed in Eq. (16), $T_{\text{eff}}(\boldsymbol{k}) \sim |\boldsymbol{E} \cdot \boldsymbol{v}(\boldsymbol{k})| / \Gamma$, which depends on $\boldsymbol{k}$. In a




\begin{thebibliography}{*}

\bibitem{French}
Stoliar, P.; Cario, L.; Janod, E.; Corraze, B.; 
Guillot-Deudon, C.; Salmon-Bourmand, S.; Guiot, V.; Tranchant,
J.; Rozenberg, M. \textit{Adv. Mater.} \textbf{2013}, \textit{25}, 3222.

\bibitem{French2}
Guiot, V.; Cario, L.; Janod, E.; Corraze, B.; Phuoc,
V. Ta; Rozenberg, M.; Stoliar, P.; Cren, T.; Roditchev, D.
\textit{Nat. Commun.} \textbf{2013}, \textit{4}, 1722.

\bibitem{Tokura}
Kumai, R.; Okimoto, Y.; Tokura, Y.
  \textit{Science} \textbf{1999}, \textit{284}, 1645.

\bibitem{Parkin}
Jeong, J.; Aetukuri, N.; Graf, T.; Schladt, T. D.; 
Samant, M. G.; Parkin, S. S. P. \textit{Science} \textbf{2013}, \textit{339}, 1402.

\bibitem{inoue}
Inoue, I. H.;  Yasuda, S.; Akinaga, H.; Takagi, H. \textit{Phys. Rev. B} \textbf{2008}, \textit{77}, 035105.

\bibitem{jslee} 
Lee, J. S.; Lee, S.; Noh, T. W. \textit{Appl. Phys. Rev.}
{\bf 2015}, \textit{2}, 031303.

\bibitem{Doug}
  Lee, S.; Fursina, A.; Mayo, J. T.; Yavuz, C. T.; Colvin, V. L.; Sumesh
Sofin, R. G.; Shvets, I. V.; Natelson, D.  \textit{Nat. Mat.} \textbf{2007}, \textit{7}, 130.

\bibitem{noh}
 Lee, S. B.; Chae, S. C.; Chang, S. H.; Lee, J. S.; Park, S.; Jo, Y.;
Seo, S.; Kahng, B.; Noh, T. W. \textit{Appl. Phys. Lett.}
\textbf{2008}, \textit{93}, 252102.

\bibitem{shukla} 
Shukla, N.; Joshi, T.; Dasgupta, S.; Borisov, P.; Lederman, D.; Datta, S. \textit{Appl. Phys. Lett.}
{\bf 2014}, \textit{105}, 012108.

\bibitem{tsuji} 
Tsuji, N.; Oka, T.; Aoki, H.; \textit{Phys. Rev. B}
\textbf{2008}, \textit{78}, 235124.

\bibitem{joura} Joura, A. V.; Freericks, J. K.; Pruschke, T. \textit{Phys.
Rev. Lett.} {\bf 2008}, \textit{101}, 196401.

\bibitem{oka} Oka, T.; Arita, R.; Aoki, H. \textit{Phys.
Rev. Lett.} \textbf{2003}, \textit{91}, 066406.

\bibitem{mazza1} Mazza, G.; Amaricci, A.; Capone, M.; Fabrizio, M. 
\textit{Phys. Rev. B} \textbf{2015}, \textit{91}, 195124.

\bibitem{mazza2} Mazza, G.; Amaricci, A.; Capone, M.; Fabrizio, M. 
\textit{Phys. Rev. Lett.} \textbf{2016}, \textit{117}, 176401.

\bibitem{Chudnovskii} Chudnovskii, F. A.; Pergament, A. L.; 
Stefanovich, G. B.;
Metcalf, P. A.; Honig, J. M. \textit{J. Appl. Phys.} \textbf{1998}, \textit{84}, 2643.

\bibitem{Brockman}
Brockman, J. S.; Gao, L.; Hughes, B.; Rettner, C. T.; Samant, M. G.; Roche, K. P.; 
Parkin, S. S. P. \textit{Nat. Nanotechnol.} \textbf{2014}, \textit{9}, 453.

\bibitem{Guenon}
Gu\'{e}non, S.; Scharinger, S.; Wang, S.; Ramirez, J. G.; 
Koelle, D.; Kleiner, R.; Schuller, I. K. \textit{EPL Europhys. Lett.}
\textbf{2013}, \textit{101}, 57003.

\bibitem{VO2-Bae}
Bae, S.-H.; Lee, S.; Koo, H.; Lin, L.; Jo, B. H.; Park, C.; Wang, Z. L. \textit{Adv.
Mater.} \textbf{2013}, \textit{25}, 5098.

\bibitem{CRO-Nakamura}
Nakamura, F.; Sakaki, M.; Yamanaka, Y.; Tamaru, S.; 
Suzuki, T.; Maeno, Y. \textit{Sci. Rep.} \textbf{2013}, \textit{3}, 2536.

\bibitem{VO2-Joule-Basov}
 Driscoll, T.; Kim, H.-T.; Chae, B.-G.; Di Ventra, M.; Basov,
 D. N.  \textit{Appl. Phys. Lett.} \textbf{2009}, \textit{95}, 043503.

\bibitem{zimmers}
  Zimmers, A.; Aigouy, L.; Mortier, M.; Sharoni, A.; Wang, S.; West, K. G.; 
Ramirez, J. G.; Schuller, I. K. 
\textit{Phys. Rev. Lett.} \textbf{2013}, \textit{110}, 056601.

\bibitem{singh} 
Singh, S.; Horrocks, G.; Marley, P. M.; Shi, Z.; Banerjee, S.; 
 Sambandamurthy, G. \textit{Phys. Rev. B} \textbf{2015}, \textit{92}, 155121.


\bibitem{Morin59} 
Morin, F.; \textit{Phys. Rev. Lett.} \textbf{1959}, \textit{3}, 34.

\bibitem{Takei66} 
Takei, H.; Koide, S. \textit{J. Phys. Soc. Jpn.} \textbf{1966}, \textit{21}, 1010.

\bibitem{duchene} 
Duchene, J.; Terraillon, M.; Pailly, P.; Adam, G. \textit{Appl. Phys. Lett.} {\bf 1971}, \textit{19}, 115.

\bibitem{berglund} 
Berglund, C. N. \textit{IEEE Trans. Elec. Dev.} {\bf 1969}, \textit{16}, 432.

\bibitem{ridley} 
Ridley, B. K. \textit{Proc. Phys. Soc.} {\bf 1963}, \textit{82}, 954.

\bibitem{CopePenn68} 
Cope R. G.; Penn, A. W.; \textit{Brit. J. Appl. Phys. (J.  Phys. D)} \textbf{1968}, \textit{1}, 161.

\bibitem{hyuntakkim} 
Kim, H.-T.; Kim, B.-J.; Choi, S.; Chae, B.-G.; Lee, Y. W.; Driscoll, T.; Qazilbash, M. M.; Basov, D. N. 
\textit{J. Appl. Phys.} {\bf 2010}, \textit{107}, 023702.

\bibitem{VO2-JKim} 
Kim, J.; Ko, C.; Frenzel, Alex.; Ramanathan, S.; Hoffman, J. E. \textit{Appl. Phys. Lett.} \textbf{2010}, \textit{96}, 213106.


\bibitem{Frenchreview} 
Janod, E.; Tranchant, J.; Corraze, B.; 
Querr\'{e}, M.; Stoliar, P.; Rozenberg, M.; Cren, T.; Roditchev, D.; Ta
Phuoc, V.; Besland, M.-P.; Cario, L. \textit{Adv. Funct. Mater.} \textbf{2015},
\textit{25}, 6287.

\bibitem{driscoll} 
Driscoll, T.; Quinn, J.; Di Ventra, M.; Basov, D. N.; Seo, G.; Lee, Y. W.; Kim, H.-T.; 
Smith, D. R. \textit{Phys. Rev. B} {\bf 2012}, \textit{86}, 094203.

\bibitem{rozenberg2014} 
Stoliar, P.; Rozenberg, M.; Janod, E.; Corraze, B.; 
Tranchant, J.; Cario, L. \textit{Phys. Rev. B} {\bf 2014}, \textit{90}, 045146.

\bibitem{garland} 
Dubson, M. A.; Hui, Y. C.; Weissman, M. B.; 
Garland, J. C. \textit{Phys. Rev. B} \textbf{1989}, \textit{39}, 6807.

\bibitem{potthoff}
Potthoff, M.; Nolting, W. \textit{Phys. Rev. B} \textbf{1999},
\textit{60}, 7834.

\bibitem{dobro}
Dobrosavljevi\'c, V.; Kotliar, G. \textit{Phys. Rev. Lett}
\textbf{1997}, \textit{78}, 3943.

\bibitem{prl15}
Li, J.; Aron, C.; Kotliar, G.; Han, J. E. \textit{Phys. Rev. Lett.}
\textbf{2015}, \textit{114}, 226403.

\bibitem{Camille2}Aron, C. \textit{Phys. Rev. B} \textbf{2012}, \textit{86}, 085127.

\bibitem{Amaricci}Amaricci, A.; Weber, C.; Capone, M.; Kotliar, G. 
\textit{Phys. Rev. B} \textbf{2012}, \textit{86}, 085110.

\bibitem{graz}
Neumayer, J.; Arrigoni, E.; Aichhorn, M.; von der
Linden, W. \textit{Phys. Rev. B} \textbf{2015}, \textit{92}, 125149.

\bibitem{satoshi} 
Okamoto, S. \textit{Phys. Rev. Lett.} \textbf{2008}, \textit{101}, 116807.

\bibitem{Camille1}
Aron, C.; Kotliar, G.; Weber, C. 
\textit{Phys. Rev. Lett.} \textbf{2012}, \textit{108}, 086401.

\bibitem{rubtsov}
Ribeiro, P.; Antipov, A. E.; Rubtsov, A. N. \textit{Phys.
Rev. B} \textbf{2016}, \textit{93}, 144305.

\bibitem{han2}
Han, J. E.; Li, J. \textit{Phys. Rev. B} \textbf{2013}, \textit{88}, 075113. 

\bibitem{han1}
Han, J. E. \textit{Phys. Rev. B} \textbf{2013}, \textit{87}, 085119.

\bibitem{Millis}
Mitra A.; Millis, A. J. 
\textit{Phys. Rev. B} \textbf{2008}, \textit{77}, 220404(R).

\bibitem{teff} 
Without the bulk dissipation, the effective temperature
inside the sample
away from the leads reaches the energy scale $\sim EL=V_{\rm s}$, unrealistically high
value. Furthermore, the RS is controlled by $E$, not by $EL$.

\bibitem{suppl} See the Supporting Information.

\bibitem{slater} Slater, J. C. \textit{Phys. Rev.} {\bf 1951}, \textit{82}, 538.

\bibitem{sugimoto}Sugimoto, N.; Onoda, S.; Nagaosa, N. 
\textit{Phys. Rev. B} \textbf{2008}, \textit{78}, 155104.

\bibitem{LZ} Zener, C. \textit{Proc. R. Soc. Lond. A} {\bf1932}, \textit{137}, 696.

\bibitem{percolation}Reynolds, P. J. ; Stanley, H. E.; Klein, W. \textit{Phys.
Rev. B} {\bf 1980}, \textit{21}, 1223.
\end{thebibliography}
\end{document}